 \documentclass[final,5p,times,twocolumn]{elsarticle}

\usepackage{amssymb}
\usepackage{lipsum}
\usepackage[breaklinks=true,colorlinks=true,citecolor=blue,linkcolor=magenta]{hyperref}
\biboptions{numbers,sort&compress}

\begin{document}

\begin{frontmatter}

\title{Full-Spectrum Quantum Simulation for the Nuclear Shell Model}

\author[first]{B. Maheshwari}
\author[second]{P. Stevenson}
\author[third]{P. Van Isacker}
\affiliation[first]{organization={Variable Energy Cyclotron Centre}, address={1/AF, Bidhan Nagar}, pin={700064}, city={Kolkata}, country={India}}
\affiliation[second]{organization={School of Mathematics and Physics, University of Surrey}, address={Guildford}, pin={GU2 7XH}, city={Surrey}, country={United Kingdom}}
\affiliation[third]{organization={Grand Accélérateur National d'Ions Lourds}, address={CEA/DSM-CNRS/IN2P3}, street={Bvd Henri Becquerel}, pin={F-14076}, city={ Caen}, country={France}}

\begin{abstract}

The nuclear shell model is a general way of expressing the many-body nuclear Hamiltonian and deciphering the underlying nuclear structure. In today’s era of modern and high-power computation, the primary limitation of the nuclear shell model is the enormous dimensionality of its Hilbert space, which far exceeds available storage capacity and prevents the diagonalization of the full Hamiltonian matrix in that space. Quantum computing offers a scalable solution to bypass this curse of dimensionality. In this work, we introduce a single-run quantum simulation capable of obtaining multiple shell-model eigenstates simultaneously. The nuclear Hamiltonian is transformed from a bit to a qubit basis using the Jordan-Wigner transformation, explicitly preserving fermionic anti-commutation. We employ a Subspace Search Variational Quantum Eigensolver (SSVQE) along with an Adaptive Derivative-Assembled Pseudo-Trotter (ADAPT) ansatz to construct the quantum circuit required to solve the shell-model problem. The ADAPT-SSVQE algorithm uses a symmetry-preserving single and double-excitation operator pool and optimizes a weighted energy sum to obtain the simultaneous convergence of all eigenstates within a targeted $M_J$ subspace, eliminating the need for post-processing efforts to extract excited spectra. We benchmark this approach by solving the problem for two and three identical nucleons in a $j=9/2$ orbital, successfully extracting five and ten mutually orthogonal states, respectively, within a 10-qubit active space. The algorithm achieves spectroscopic accuracy, in simulation, relative to exact diagonalization and intrinsically restores total angular momentum $(\hat{J}^2)$ symmetry.
\end{abstract}

\begin{keyword}
Nuclear Shell Model \sep Quantum Simulation \sep ADAPT-SSVQE 
\end{keyword}
\end{frontmatter}

\section{Introduction}

Atomic nuclei are one of the most complex quantum many-body systems. They are comprised of nucleons, that is, protons and neutrons, held together by the strong nuclear force. Among many successful many-body methods describing the structure of nuclei, the nuclear shell model is a standard and general paradigm for giving a full quantum description in general, while allowing specific approximations. In this model, nucleons move in a mean-field potential augmented with a residual interaction that induces many-body correlations. Diagonalizing the Hamiltonian in the single-particle mean-field basis, one obtains the energies of nuclear states~\cite{smbook}. Even in today's era of modern and high-power computation, the exact diagonalization of shell-model Hamiltonians rapidly becomes intractable for large model spaces due to exponential scaling of many-body Hilbert space. To circumvent the limitations caused by dimensionality, quantum computing can offer a scalable approach based on the principles of superposition and interference of qubit states. 

The quantum simulation of complex many-body systems is at the frontier of quantum computing with many potential applications in quantum chemistry, condensed-matter physics, nuclear physics, high-energy physics, among others. The quantum algorithms and hardware are now advancing rapidly providing hope to solve problems that otherwise remain formidable. So far, quantum algorithms such as the variational quantum eigensolvers (VQE)~\cite{vqe} are among the most successful ones, offering approximate solutions for many-body problems of physics and chemistry on near-term quantum devices~\cite{nisq}. This success crucially depends on the choice of ansatz that controls the evolution of states by designing the quantum circuit responsible for the allowed fermionic correlations and interactions~\cite{lacroix2020,guzman2022,guzman2024,gibbs2025}. Two commonly used methods for ansatz preparation are: Unitary-Coupled Cluster (UCC)~\cite{ucc} variants; and Adaptive Derivative-Assembled Pseudo Trotter Variational Quantum Eigensolver (ADAPT-VQE)~\cite{adapt}. While a standard unitary coupled cluster ansatz~\cite{ucc} is general but computationally expensive, adaptive approaches~\cite{adapt} construct an ansatz iteratively ensuring compactness of representation. 

The quantum simulation of ground-state solutions for the nuclear shell model has recently been attempted using ADAPT-VQE~\cite{spain}. A more specific quantum simulation of the shell model applied to $^{58}$Ni has been demonstrated using a problem-specific ansatz optimizing separately for the ground and two excited states~\cite{bharti}. Some other example works include the solution of the pairing problem~\cite{jiang2023,zhang2024,zhang2025,liu2025,acta}. However, simulating the full excited spectra of atomic nuclei remains an interesting and challenging endeavour. The desire for complete spectroscopy - knowledge of all energy levels - is a fundamental concern in nuclear physics whereas it is perhaps not as critical in many electronic systems of interest. Since nuclei undergo radioactive decay to excited states of daughter products, which then subsequently decay through gamma-ray cascades, detailed knowledge of excited levels is necessary for our complete understanding and also for applications in areas such as nuclear reactor decay heat \cite{algora}. An initial application of ADAPT-VQE with subspace search (SSVQE~\cite{ssvqe}) to simulate a pairing Hamiltonian for a simple four-qubit system was presented in a conference~\cite{acta}. In this work, we present the full details of method along with the example system of ten qubits within the framework of the nuclear shell model for the first time. We further demonstrate the success of the designed algorithm for calculating the realistic low-energy spectra dominated by the single-$j$ $0g_{9/2}$ orbital. This approach enables the simultaneous approximation of mutually orthogonal ground and excited states while respecting angular momentum symmetries in a single optimization run. 

\section{Algorithm}

In the effective nuclear shell model, the Hamiltonian $H_{eff}$ is typically truncated to one- and two-body terms, expressed in second quantization~\cite{smbook}:
\begin{equation}
    H_{eff}=\sum_{ab} \epsilon_{ab} a^\dagger_a a_b + \frac{1}{4} \sum_{abcd} \langle ab | V| cd \rangle_A a^\dagger_a a^\dagger_b a_d a_c 
\end{equation}
with $a^\dagger_i$ and $a_i$ being the creation and annihilation operators for a single-particle orbital $i$ in the $m$-scheme, $\epsilon_{ab}$ represent single-particle energies. 
Two-body matrix elements, $\langle ab |V|cd\rangle_A $ describe the interaction between the anti-symmetrized states. These elements are converted from a $jj$-coupled basis with defined total angular momentum into an uncoupled $m$-scheme basis. The computational challenge of the nuclear shell model lies in the size of the Hilbert space. For a system of two nucleons in the ten $m$-scheme states of the $0g_{9/2}$ orbital, the dimension is manageable $^{10}C_2= 45$. However, as we move toward systems with a larger $j$, or toward multi-$j$ orbitals, the complexity scales rapidly. Larger angular momentum $j$ or multiple $j$-orbitals increase the number of $m$-scheme qubits, $(2j+1)$ or $\sum(2j+1)$. For many-nucleon systems in a large valence space, the Hilbert space dimensions can easily exceed $\sim{10}^{10}$, such that classical diagonalization becomes impossible. This necessitates the development of quantum-based algorithms to overcome this challenge.  

To map the fermionic Hamiltonian onto a quantum computer, we use the Jordan-Wigner (JW) transformation~\cite{jw}. This procedure translates creation and annihilation operators for the $j^{\mathrm{th}}$-orbital into qubit-based Pauli operators as follows:
\begin{eqnarray}
    a_j \rightarrow \Big[ \Pi_{k=1}^{j-1} \mathrm{Z}_k \Big] \frac{1}{2} (\mathrm{X}_j+ i \mathrm{Y}_j) \\
    a^\dagger_j \rightarrow \Big[ \Pi_{k=1}^{j-1} \mathrm{Z}_k \Big] \frac{1}{2} (\mathrm{X}_j- i \mathrm{Y}_j)   
\end{eqnarray}
where X, Y, Z are Pauli spin matrices. The Z-string in square brackets is critical; it preserves the fermionic anti-communication across different qubits, ensuring that the anti-symmetry of the many-body wavefunction is maintained. Once transformed, the Hamiltonian is represented as a sum of Pauli strings, a format directly measurable on quantum hardware. Using the principles of superposition and entanglement, a quantum computer with $N$ qubits can naturally represent a state in a Hilbert space of dimension $2^N$.  Hence a Hilbert space of dimension $10^{10}$, around the limit of current classical calculation, can be represented on $N=\lceil{10/\log_{10}2}\rceil=34 $ qubits.

The VQE~\cite{vqe} is a hybrid quantum-classical algorithm designed for Noisy Intermediate Scale Quantum (NISQ) devices~\cite{nisq}. It relies on the variational principle that the expectation value of a trial wavefunction is always an upper bound to the true ground-state energy. A parameterized trial state is chosen to be $|\Psi(\vec\theta) \rangle = U(\vec\theta) | \phi_0 \rangle$ using a quantum circuit. The energy expectation value $\langle \Psi(\vec\theta) | \hat{H} | \Psi(\vec\theta) \rangle$ is measured on the quantum computer, and the parameters $\vec\theta$ are optimized iteratively by a classical optimizer, leading to energy minimization. The ansatz, in the form of a quantum circuit, typically consists of single-qubit rotation gates $(R_x(\theta),R_y(\theta),R_z(\theta))$ and multi-qubit entangling gates such as CNOT. These are the building blocks for imposing essential physical correlations between nucleons. Specific quantum computing hardware may compile a circuit to different native gate sets. One-qubit plus CNOT gates are a typical way of describing NISQ-era circuits, with the CNOT count an inverse measure of utility, as the two-qubit CNOT gates are the most error-prone on current hardware~\cite{wong}.      

While the standard VQE targets the ground state, Subspace Search Variational Quantum Eigensolver (SSVQE)~\cite{ssvqe} extends this capability to find ground and multiple excited states simultaneously. This is achieved by evolving a set of $k$ mutually orthogonal states $|\phi_i\rangle$ through the same unified ansatz $U(\vec\theta)$. The algorithm minimizes a weighted loss function defined by 
\begin{eqnarray}
L(\vec\theta) = \sum_{i=0}^{k-1} w_i \langle \phi_i | U^\dagger (\vec\theta) \hat{H} U(\vec\theta) | \phi_i \rangle  = \sum_{i=0}^{k-1} w_i E_{i} (\vec\theta)
\end{eqnarray}
where $w_i$ are descending positive weights adopted as $w_i=k-i$. 
By assigning the highest weight to the lowest index, the optimizer prioritizes the convergence of the ground state, followed by the first excited state, and so on. 

The version of SSQVE used here, ADAPT-SSQVE, uses the ADAPT-VQE idea of dynamically constructing the ansatz $U(\vec\theta)$ by iteratively selecting the most useful operators from a predefined pool $P$. This ensures the resulting quantum circuit remains minimally complex while effectively capturing essential nucleonic correlations. The calculations are restricted to a fixed $M_J=\sum_i m_i$ subspace, since the total magnetic projection $M$ is a good quantum number of the Hamiltonian, and $M$-conservation is naturally accommodated in the $m$-scheme approach. Specifically, $M$-conservation is enforced by a symmetry-constrained pool consisting of double-excitation operators, $T\equiv T_{ab}^{cd}=a_c^\dagger a_d^\dagger a_b a_a$, where the magnetic projection indices are constrained such that ${m_j}_c+{m_j}_d={m_j}_a+{m_j}_b$. These fermionic operators are transformed into Hermitian Pauli operator $A \in P$ suitable for an exponential ansatz, $\exp(A)$, defined by $A=i(T-T^\dagger)$ where $T^\dagger$ is the complex conjugate of $T$. 
By restricting the pool to the $M_J$-conserving excitations, the physical relevance and compactness of the quantum circuit can be ensured.

At each iteration $iter$, the next operator $A_{iter}$ is chosen from pool of operators $P$ by maximizing the weighted absolute gradient contribution $G_{iter}$ to the SSVQE loss function,
\begin{equation}
    G_{iter}={max}_{A\in P} \Bigg| \sum_{i=0}^{k-1} w_i \langle \Psi_i(\vec\theta) | i [H, A_{iter}] | \Psi_i(\vec\theta) \rangle \Bigg| 
\end{equation}
The ADAPT algorithm chooses the operator with maximum gradient to evolve the ansatz and append it to the circuit. The expanded parameter vector $\theta_n=(\theta_{n-1},\theta_{new})$, is then fully re-optimized using the L-BFGS-B algorithm~\cite{lbfgs}. To guarantee monotonic improvement in the energies, the optimizer is initialized with the converged parameters of the previous step, setting the newly appended parameter to zero, $\theta_{new}=0$.  
This iterative evolution proceeds until $G_{iter}$ falls below a defined tolerance, ${10}^{-2}$ in this simulation, or when the energy variance between consecutive iteration satisfies $\Delta E < {10}^{-2}$. The final ADAPT ansatz is a product of Pauli evolution gates,
\begin{equation}
    U(\vec\theta)=\Pi_{iter=0}^{iter_{max}}exp(-i \vec\theta_{iter} A_{iter})
\end{equation}
which is unitary, so if the initial states are chosen to be mutual orthogonal then the evolution using $U(\vec\theta)$ will strictly maintain their orthogonality and symmetries without any post-efforts.

Because the simulation operates strictly within an $M_J$ subspace, the wavefunctions are not guaranteed to be exact eigenstates of the total angular momentum operator $\hat{J}^2$ prior to convergence. The restoration and verification of angular momentum symmetry are crucial for nuclear spin and parity assignment. We explicitly construct the $\hat{J}^2$ operator in second quantization, $\hat{J}^2=\hat{J}_z^2 + \frac{1}{2} (\hat{J}_+ \hat{J}_- + \hat{J}_- \hat{J}_+)$ to verify the physical nature of the generated spectra. The ladder operators $\hat{J_{+}}$ and $\hat{J_{-}}$ are generated using Clebsch-Gordan coefficients corresponding to the $0g_{9/2}$ orbital. Upon convergence of ADAPT-SSVQE algorithm, these operators are mapped to the qubit basis, and the expectation value $\langle \hat{J}^2 \rangle$ for each orthogonal state is measured. By solving $\langle \hat{J}^2 \rangle=J(J+1)$, we assign the $J$ to the resulting states, confirming that the dynamic ansatz successfully captures the targeted nuclear symmetries which are not specifically enforced by construction.   

\begin{figure}
    \centering
    \includegraphics[width=1.0\linewidth]{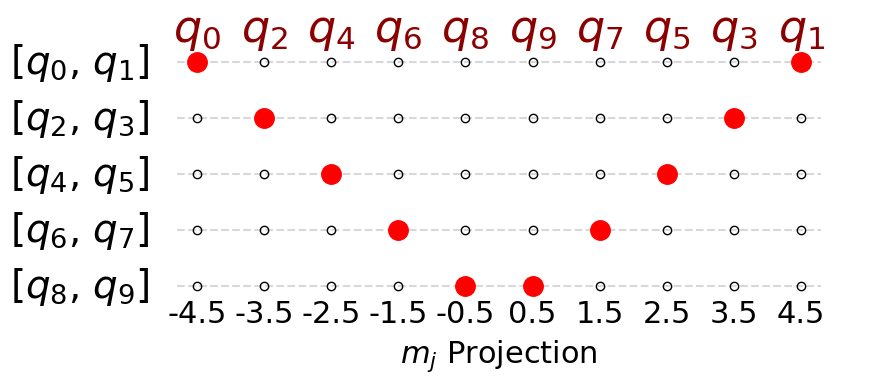}
    \caption{A 10-qubit mapping and initial orthogonal Hartree-Fock (HF) states used in the current quantum simulation for two nucleons in $0g_{9/2}$ orbital. Red circles denote the occupied single-particle states in m-scheme.}
    \label{fig:map}
\end{figure}

\section{Results and Discussion}

\begin{figure*}
    \centering
\includegraphics[width=0.85\linewidth]{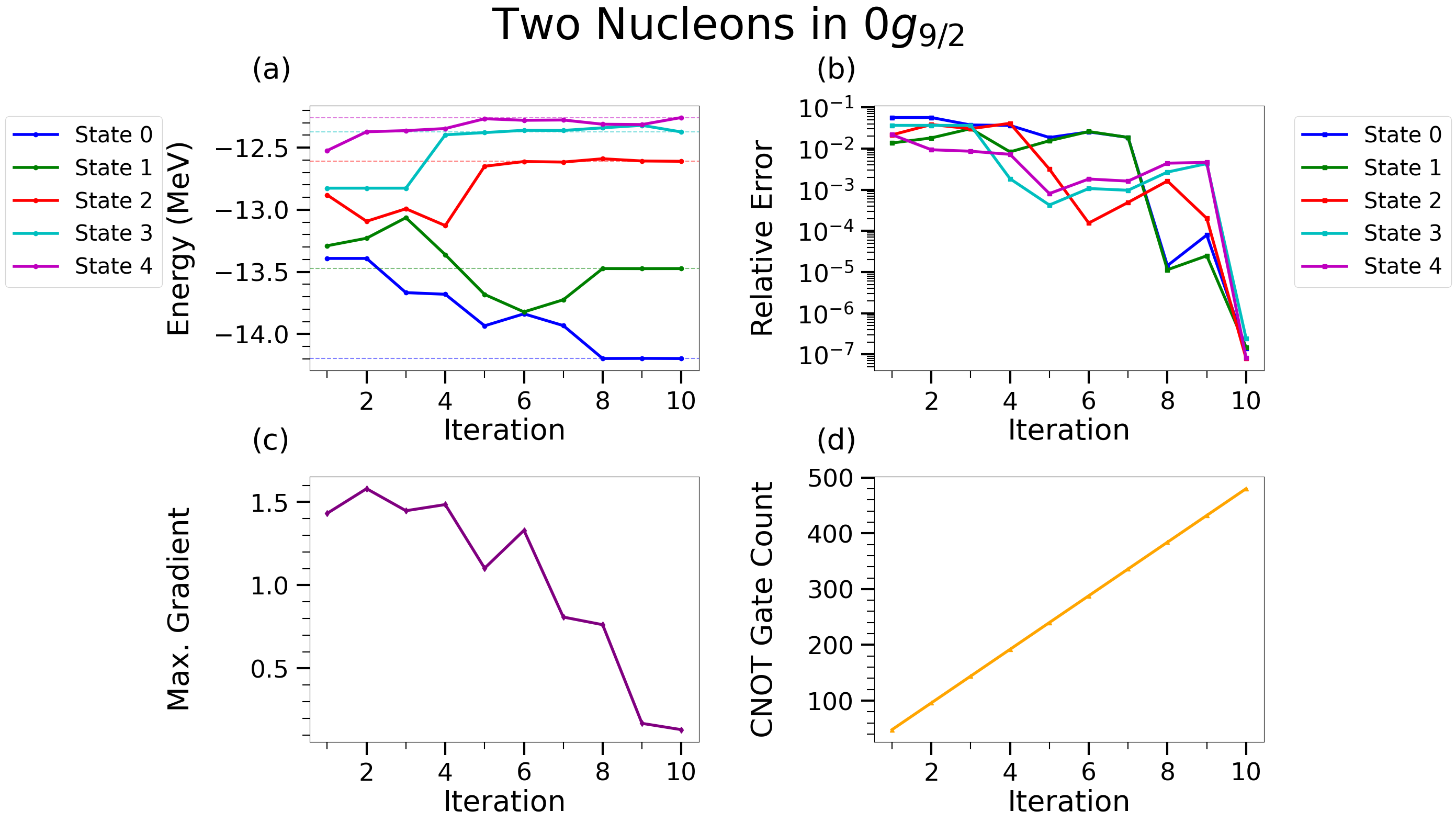}
    \caption{Variation of (a) simulated energies (in MeV) for ground and excited states shown in different color along with exact solutions shown by dashed lines (b) relative error defined by (simulated - exact) /exact (c) maximum gradient (in MeV) of the appended operator (d) CNOT gate count, across iterations for two nucleons in $0g_{9/2}$ orbital. }
    \label{fig:g92}
\end{figure*}

To benchmark the ADAPT-SSVQE algorithm, we have used exact state vector simulation to study two- and three-nucleon systems occupying a 10-qubit $0g_{9/2}$ orbital,  as shown in Fig.~\ref{fig:map} for the two-nucleon case. The single-particle energies and two-body matrix elements are adopted from the JUN45 interaction~\cite{jun45}.  

For the two-nucleon case,  the simulation is constrained to the $M_J=0$ subspace and utilized $M_J=0$ conserving operators from the total operator pool of $M_J$-conserving double excitation operators in $0g_{9/2}$ orbital. To monitor angular momentum restoration, a $\langle \hat{J}^2 \rangle$ operator consisting of 270 terms is also explicitly constructed via the JW mapping. The algorithm initializes by acting on five mutually orthogonal Hartree-Fock basis states representing the valid $m_j$ pairs, $[q_0,q_1]$, $[q_2,q_3]$, $[q_4,q_5]$, $[q_6,q_7]$, and $[q_8,q_9]$ as depicted in Fig.~\ref{fig:map}. Fig.~\ref{fig:g92}(a) shows the step-by-step energy convergence of these targeted states approaching the exact solutions which are indicated by dashed horizontal lines. The high precision of this approach is emphasized in Fig. 2(b) which shows the relative error dropping  below ${10}^{-6}$ for all five states by the final iteration. 

\begin{table*}[!htb]
    \centering
    \caption{State vector composition of the five states for two nucleons in the $0g_{9/2}$ orbital in qubit basis after the ADAPT-SSVQE convergence. }
    \begin{tabular}{|c|c|c|c|}
       \hline
       State Number  & State vector Composition & $\langle \hat{J}^2 \rangle$ & $J$\\
       \hline
       State 0 & $ 0.4472[q_4, q_5]+0.4472[q_0, q_1]-0.4472 [q_6, q_7]-0.4472[q_2, q_3]+0.4472 [q_8, q_9]$ & 0.0 & 0\\
       State 1 & $ 0.7385 [q_0, q_1]-0.4924 [q_8, q_9]+0.3693 [q_6, q_7]-0.2462 [q_2, q_3]-0.1231 [q_4, q_5]$ & 6.0 & 2 \\
       State 2 & $  0.5818 [q_2, q_3]+0.4760 [q_0, q_1]+0.4760 [q_8, q_9]-0.4495 [q_4, q_5]-0.0793 [q_6, q_7] $ & 20.0 & 4\\
       State 3 & $ -0.6055 [q_2, q_3]-0.5505 [q_4, q_5]+0.4404 [q_8, q_9]+0.3303 [q_6, q_7]-0.1651 [q_0, q_1]$ & 42.0 & 6\\
       State 4 & $ -0.7404 [q_6, q_7]-0.5289 [q_4, q_5]-0.3702 [q_8, q_9]-0.1851 [q_2, q_3]-0.0264 [q_0, q_1]$ & 72.0 & 8\\
       \hline
    \end{tabular}
    \label{tab:2n}
\end{table*}

During the adaptive iterations, the ansatz dynamically grows by selecting the operator with the steepest gradient as detailed in Fig.~\ref{fig:g92}(c). For this system, the algorithm selects an initial operator driven by a high initial gradient of $G_1=1.432$ MeV. The classical L-BFGS-B optimizer dynamically adjusts the corresponding rotational parameter for this operator, converging at a value of $\theta_1=-0.967$ to successfully minimize the initial energy manifold. The selected $M_J$-conserving double-excitation operator is mapped via JW transformation expanding into Pauli strings, $A_{Pauli}$. Enforcing Hermiticity cancels all terms containing an even number of Pauli $\mathrm{Y}$ operators. The eight surviving strings contain exactly one or three $\mathrm{Y}$ operators, along with a $\mathrm{Z}$-string representing the intermediate qubits to preserve fermionic anti-commutation. The sign of each term is strictly governed by the parity of $\mathrm{Y}$ matrices: terms where the majority of $\mathrm{Y}$ operators act on the creation indices are positive while those dominated by $\mathrm{Y}$ operators on the annihilation indices are negative. The execution of the exponentiated operator, $\exp(-i\theta A_{Pauli})$ requires the sequential simulation of all eight strings using a CNOT staircase. For each four-qubit interaction, this involves three CNOT gates entangling the qubits to isolate their combined state on a target qubit. This is followed by a single rotation gate applying the optimized $\theta$ parameter. Finally, the circuit must reverse the CNOT staircase to unentangle the qubits using three additional CNOT gates to restore the state orthogonality. This configuration results in a total cost of six CNOT gates per Pauli string yielding a baseline total of 48 CNOT gates for the first double-excitation operator.   

As the iterations progress, the optimizer continuously extracts the correlation energy. By iteration 5, the ground state energy drops significantly to $-13.933$ MeV using 240 CNOT gates. The algorithm seamlessly halts at iteration 11 when the maximum gradient plummeted below the convergence threshold. This indicates that the operator pool's correlation capacity is fully exhausted. The final unified ansatz optimizes a total of 10 parameters and accumulates a circuit complexity of 480 CNOT gates, that is 48 CNOT gates per iteration for each selected multi-term Pauli operator, as shown in Fig.~\ref{fig:g92}(d). This happens when the Pauli $\mathrm{Z}$-string completely cancels out due to the fully-coupled orbitals involved in the excitation.  
Ultimately, the algorithm achieves perfect fidelity relative to the classical exact diagonalization within 10 iterations. Across the entire five-state spectrum, the residual error (difference between simulated and exact energies) is reduced to zero. The optimized set of energies for the five states converged to exactly $-14.197$, $-13.473$, $-12.610$, $-12.373$, $-12.257$ MeV, respectively.  

A critical challenge in variational quantum simulations of nuclear structure is the preservation of fundamental symmetries, particularly the total angular momentum $\hat{J}^2$. 
While our ADAPT operator pool is designed to strictly conserve particle number and $z$-projection of angular momentum $(M_J)$, its members do not explicitly commute with the $\hat{J}^2$ operator. Despite this, beyond energy eigenvalues, the simulation successfully restores the rotational symmetry of the nucleus. Post-convergence evaluation of the $\langle \hat{J}^2 \rangle$ operator yields precise integer assignments for all five states, perfectly matching the allowed states of the $0g_{9/2}$ orbital as $J=0,2,4,6,8$. This confirms that the $M_J$-conserving pool of operators intrinsically preserve full SU(2) rotational symmetry within the target manifold. Furthermore, an analysis of final state vectors provides direct insight into the configuration-mixing governed by the angular momentum coupling of two valence nucleons. The wave function for the $J=0$ ground state evolves into a perfectly balanced superposition. All five $m_j$ pairs $[q_0,q_1]$, $[q_2,q_3]$, $[q_4,q_5]$, $[q_6,q_7]$, and $[q_8,q_9]$ contribute with an identical absolute amplitude of 0.4472, representing a $\frac{1}{5}$ probability each. This matches the theoretical benchmark of a pure $J=0$ pairing state in a single-j orbital, confirming that the algorithm successfully reconstructs the pairing force. Every state vector component shown in Table~\ref{tab:2n} can be identified, up to an overall sign for each state, with the Clebsch-Gordan coefficient $(jm_j \, j-m_j|J0)$ multiplied by the statistical factor $\sqrt{2}$. The components are therefore geometric quantities, independent of the interaction between the nucleons.

\begin{figure*}
    \centering
\includegraphics[width=0.85\linewidth]{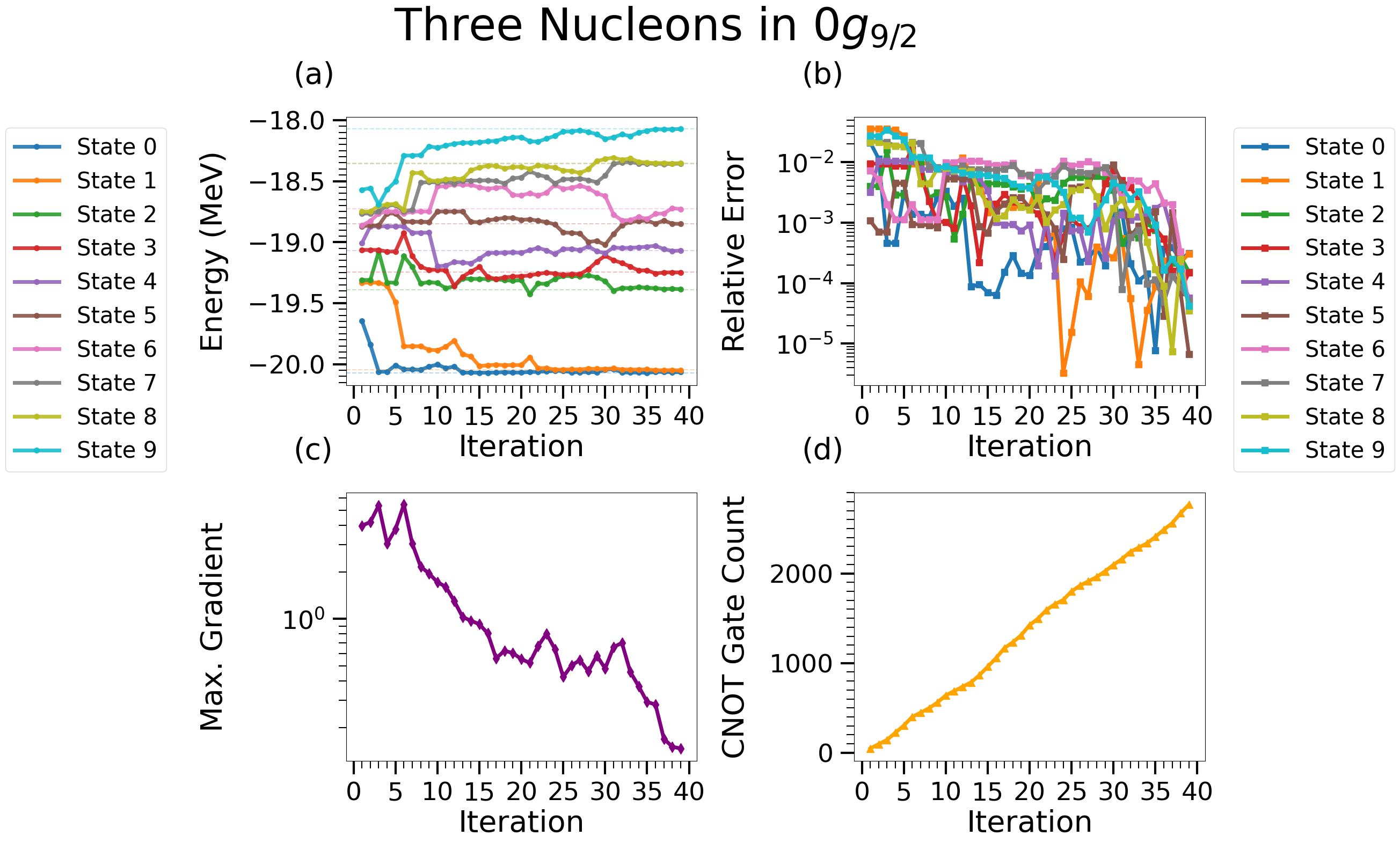}
    \caption{Variation of (a) energies (in MeV) for ground and excited states shown in different color along with exact solutions shown by dashed lines (b) relative error (c) maximum gradient (in MeV) of the appended operator (d) CNOT gate count, across iterations for three nucleons in $0g_{9/2}$ orbital.}
    \label{fig:g93}
\end{figure*} 

To evaluate the stability of algorithm, and applicability to more realistic cases, we investigate a three-nucleon system in the $0g_{9/2}$ orbital using the realistic JUN45 shell model interaction~\cite{jun45}.  We target the 10 mutually orthogonal states within the $M_J=0.5$ subspace. The simulation is initialized using the allowed mutually orthogonal Hartree-Fock states: $[q_0, q_1, q_9]$, $[q_2, q_3, q_9]$, $[q_4, q_5, q_9]$, $[q_6, q_7, q_9]$, $[q_0, q_3, q_7]$, $[q_1, q_2, q_8]$, $[q_1, q_4, q_6]$, $[q_3, q_4, q_8]$, $[q_2, q_5, q_7]$, $[q_5, q_6, q_8]$. During the first iteration, the ADAPT-SSVQE algorithm identifies the best operator with a maximum gradient of $G_1=3.978$ MeV and optimize a single parameter to $\theta_1=0.591$. This yields an initial energy spectrum of $-19.645$, $-19.331$, $-19.309$, $-19.063$, $-19.008$, $-18.870$, $-18.863$, $-18.765$, $-18.748$, $-18.570$ MeV. As the ansatz expands, the spectrum systematically refines, as shown in Fig.~\ref{fig:g93}. The energy convergence demonstrates the steady optimization of the target manifold; by the tenth iteration, 10 parameters are optimized (0.829, 1.335, 1.019, 1.078,  0.616, $-1.324$, 0.774, 0.518, 0.304, 0.824) with an energy spectrum shown in Fig.~\ref{fig:g93}(a). The maximum gradient undergoes a characteristic non-monotonic decay as new operators are introduced to resolve specific states, falling to $G_{20}=0.551$ MeV by the twentieth iteration where 20 parameters $(\theta_1,...\theta_{20})$ (0.822, 1.371, 1.811, 1.135, 0.605, $-1.315$, 0.690, 0.542, 0.614, 0.847, $-0.105$, 0.114, $-0.923$, 0.385, 0.810, $-0.572$, $-0.305$, 0.216, $-0.193$, 0.162) stabilize the circuit. The optimization process reaches its convergence in 39 iterations, plotted in Fig.~\ref{fig:g93}(a), terminating when the maximum energy variation across all 10 evaluated states falls to roughly 0.006 MeV. As illustrated in Fig.~\ref{fig:g93}(b), despite early oscillatory behavior inherent to navigating a complex energy landscape, the relative errors reliably plunge into a  precise regime relative to exact diagonalization. Fig.~\ref{fig:g93}(c) depicts the variation of maximum gradient with each iteration. Consequently, the complexity of the quantum circuit grows linearly as shown in Fig.~\ref{fig:g93}(d); the CNOT gate count - a primary metric for quantum hardware noise and execution latency - scales from 48 CNOTs at the first iteration to a final depth of 2768 CNOTs. Ultimately, the algorithm successfully resolves 10 distinct orthogonal states, yielding final converged energies in robust agreement with the exact diagonalization results, as illustrated in Fig.~\ref{fig:g93}(a). These computed solutions highlight the efficacy of adaptive subspace-search methods for multi-nucleon simulations. Future work is required to mitigate the hardware noise due to growing CNOT gates while resolving the complex multi-nucleon spectrum in order to apply the method to real NISQ hardware.

\begin{table*}[!htb]
    \centering
    \caption{State vector composition of the ten states for three nucleons in the $0g_{9/2}$ orbital in qubit basis after the ADAPT-SSVQE convergence. Only the five largest components for each state are listed. The last column gives the nearest exact half-integer $J$ assignment corresponding to simulated $\langle \hat{J}^2 \rangle$ values.}
    \begin{tabular}{|c|c|c|c|}
       \hline
       State No.  & State vector Composition & $\langle \hat{J}^2 \rangle$ & $J$ \\
       \hline
       State 0 & $0.6156 [q_6, q_7, q_9] + 0.4115 [q_2, q_3, q_9] -0.4049 [q_0, q_1, q_9] +0.3478 [q_5, q_6, q_8] -0.3116 [q_4, q_5, q_9]$ & 22.59 & 4.5 \\ 
       State 1 & $ -0.6170 [q_5, q_6, q_8] -0.4739 [q_4, q_5, q_9] + 0.3408 [q_2, q_5, q_7] -0.3142 [q_0, q_1, q_9]+ 0.2612 [q_2, q_3, q_9] $ & 17.90 & 3.5 \\ 
       State 2 & $ 0.5191 [q_2, q_5, q_7]+ 0.5055 [q_0, q_1, q_9] -0.3364 [q_4, q_5, q_9]+ 0.3242 [q_6, q_7, q_9] -0.3149 [q_0, q_3, q_7]$ & 9.59 & 2.5 \\ 
       State 3 & $0.4757 [q_6, q_7, q_9]+ 0.4378 [q_0, q_1, q_9]-0.3932 [q_1, q_2, q_8]+0.3883 [q_3, q_4, q_8]+0.3213 [q_0, q_3, q_7] $ & 47.90  & 6.5 \\ 
       State 4 & $  0.5235 [q_0, q_3, q_7]-0.4612 [q_3, q_4, q_8]+0.4078 [q_1, q_2, q_8]-0.3285 [q_5, q_6, q_8]-0.3108 [q_4, q_5, q_9]$ & 35.61  & 5.5 \\
       State 5 & $ 0.6548 [q_1, q_4, q_6]-0.4826 [q_1, q_2, q_8]-0.3375 [q_2, q_3, q_9]-0.2853 [q_3, q_4, q_8]+0.2134 [q_5, q_6, q_8]$ & 4.00  & 1.5 \\
       State 6 & $-0.5596 [q_1, q_4, q_6]+0.4642[q_0, q_3, q_7]-0.4381 [q_2, q_3, q_9]+0.3773[q_2, q_5, q_7]-0.3640 [q_0, q_1, q_9] $ & 24.69  & 4.5 \\
       State 7 & $-0.4914 [q_1, q_2, q_8]-0.4313 [q_6, q_7, q_9]-0.3755 [q_4, q_5, q_9]+0.3263 [q_5, q_6, q_8]+0.3246 [q_2, q_3, q_9] $ & 79.22 & 8.5 \\
       State 8 & $0.4874 [q_0, q_3, q_7]+0.4563 [q_2, q_5, q_7]+0.4268 [q_3, q_4, q_8]+0.3364 [q_2, q_3, q_9]+0.3086 [q_1, q_4, q_6] $ & 65.33 & 7.5 \\
       State 9 & $0.5533 [q_4, q_5, q_9]-0.4553 [q_3, q_4, q_8]+0.4238 [q_2, q_3, q_9]-0.3282 [q_5, q_6, q_8]+0.2405 [q_6, q_7, q_9]$ & 120.64 & 10.5 \\
       \hline
    \end{tabular}
    \label{tab:3n}
\end{table*}

The extracted properties in terms of $\langle \hat{J}^2 \rangle$ and state vectors provide insight into the correlation dynamics and the preservation of rotational symmetry within the three-nucleon $0g_{9/2}$ system. For an odd-mass system, the total angular momentum $J$ must take half-integer values with the exact expectation values corresponding to $\langle \hat{J}^2 \rangle = J(J+1)$. The simulated $\langle \hat{J}^2 \rangle$ values exhibit slight symmetry breaking compared to a perfectly isotropic system most likely due to the enforced convergence threshold leading to negligible missing physical correlations, yet they closely map onto the theoretical half-integer $J$ spectrum, as listed in Table~\ref{tab:3n}, with $\langle \hat{J}^2 \rangle$ evaluated directly from the quantum circuit utilizing the complete state vector. Unlike the simplified two-nucleon pairing state, the three-nucleon wavefunctions reveal highly complex, asymmetric configuration mixing attributed to the odd nucleon. 
Table~\ref{tab:3n} presents only the five largest state vector compositions of the total wavefunction for each nuclear state. The interpretation of the components in the three-nucleon case is considerably more intricate than it is for two nucleons. In fact, with the exception of the two $J=9/2$ states, the components can be expressed as follows:
\begin{eqnarray}
\sqrt{6}\sum_{J_2\,{\rm even}}[j^2(J_2)j|\}j^3 \upsilon J](jm_1\,jm_2|J_2m_1+m_2) \nonumber \\(J_2m_1+m_2\,jm_3|JM),
\label{eq:comp3}
\end{eqnarray}
where $[j^2(J_2)j|\}j^3\upsilon J]$ is a coefficient of fractional parentage for states characterized by the seniority quantum number $\upsilon$~\cite{shalit}. Most of the components again are geometric quantities, independent of the interaction between the nucleons. The exception concerns the two $J=9/2$ states (with seniority $\upsilon=1$ and 3, respectively), whose mixing is interaction dependent. For $J\neq9/2$ the components listed in Table~\ref{tab:3n} should therefore correspond, up to an overall sign for each state, to the expression~(\ref{eq:comp3}). This provides a sensitive test on the convergence of the wave functions after 39 iterations. It is found that deviations occur for some individual coefficients but that the overall structure of the states is well reproduced. The proposed quantum algorithm hence not only accurately predicts the multi-nucleon energy spectrum but also encodes the correlated structure of the wave functions, as imposed by the combined properties of anti-symmetry and angular momentum conservation.

\section{Conclusion}

We have developed a quantum simulation algorithm, ADAPT-SSVQE, for generating simultaneously the ground and excited shell-model states in a single-optimization run. We benchmark this approach by successfully simulating the full low-lying nuclear energy spectra for systems of two and three identical nucleons in the $0g_{9/2}$ orbital using a quantum computing framework. The nuclear shell-model Hamiltonian with JUN45 interaction is mapped onto qubits via the Jordan-Wigner transformation while preserving the necessary fermionic anti-commutation relations. The variational ADAPT-SSVQE algorithm efficiently generates the five (ten) initial Hartree-Fock states corresponding to two (three) nucleons using $M_J$-conserving double-excitation operator. We have shown for the first time both the ground and excited states of the system within a few iterations with $\theta$ parameters optimized classically in a single run using this algorithm. The adaptive construction of the ansatz ensures that the circuit depth grows only when physically justified by the energy gradient. The resulting state vectors accurately capture the complex configuration mixing governed by angular momentum coupling and confirm the intrinsic restoration of the SU(2) rotational symmetry of the total angular momentum $J$ within the targeted $M_J$ subspace. 
This work presents a clear roadmap for future quantum computing research of the many-body nuclear Hamiltonian and provides a theoretically optimal and symmetry-preserving route to calculate complex nuclear spectra in a single run. Such developed algorithms are generic and may be applied to other branches of many-body physics. 

The transition from logical simulation to physical hardware execution remains a significant challenge. Preliminary evaluation of our optimized 480-CNOT circuit for the two-nucleon system under a realistic hardware noise profile demonstrates that deep CNOT chains induce depolarizing noise. Future work is required to integrate zero error mitigation strategies. Concurrently, algorithmic developments must focus on further operator pool truncation or on the exploration of symmetry-broken, hardware-efficient ansatzes to drastically reduce the circuit depth. Future simulation work is planned to expand this framework to more complex systems involving both protons and neutrons along with isospin symmetry within larger valence spaces.  

\section*{Acknowledgments}
B. M. gratefully acknowledges the funding support from ANRF (India), RJF/2025/000092, and to the HORIZON-MSCA-2023-PF-01 project, ISOON, under grant number 101150471 at GANIL (France). She also thanks A. Navin for useful discussions. This work is also funded by UK STFC grant no. ST/Y000358/1.

\appendix

\begin{small}
\begin{verbatim}

\end{verbatim}   
\end{small}

\newpage

\end{document}


\begin{frontmatter}

\title{Supplementary Material for ``Understanding Charge Radii with Machine Learning: Discovering Physics Expressions"}

\author[first]{B. Maheshwari}
\author[first]{P. Van Isacker}
\affiliation[first]{organization={Grand Accélérateur National d'Ions Lourds}, address={CEA/DSM-CNRS/IN2P3}, street={Bvd Henri Becquerel}, pin={F-14076}, city={ Caen}, country={France}}

\end{frontmatter}

We provide supplementary details of machine-learning (ML) algorithm presented in the main manuscript. A data file consisting of outputs including predictions and extrapolations from both the ML models, Light Gradient Boosting Method (LGBM)~\cite{lgbm} and Gaussian Process Regression (GPR)~\cite{gpr}, is also provided separately.

Hyperparameter optimization for the LGBM model is performed using Optuna~\cite{optuna} to minimize the root-mean-squared-error (RMSE) over 100 trials. The search space included numer of estimators $n\_estimators (100-1000)$, a learning rate on a logarithmic scale $learning\_rate (log, 0.01-0.2)$, the complexity of individual trees by $num\_leaves (20-256)$, $max\_depth (5-20)$, $min\_child\_samples (5-100)$, $subsample (0.6-1.0)$, $colsample\_bytree (0.6-1.0)$, and regularization strategies to prevent overfitting on logarithmic scale with penalty terms, $reg\_alpha (log, 0.001-10.0)$, and $reg\_lambda (log, 0.001-10.0)$. 

\begin{table*}[!htb]
\centering
\caption{Expressions derived in each fold of symbolic regression using LGBM numerical results.  }
\resizebox{0.95\textwidth}{!}{\begin{tabular}{|c|c|c|}
\hline
Fold No. & RMSE & Expression \\  [1mm]
\hline
1 & 0.053 & $-4.480\times 10^{-5}N^2 + 0.015N + 0.736Z^{1/3} + 1.136$ \\[1mm]
2 & 0.056 & $0.509A^{1/3} + 1.686\times 10^{-6}BEA - 0.012N - 0.889(0.005CF - 0.032)(0.319CF - 8.744I + 0.217) + 1.332$ \\[1mm]
3 & 0.061 & $0.855A^{1/3} - 0.520I - 0.878(0.314CF - 1.138)(0.001N - 0.140) + 0.586$ \\[1mm]
4 & 0.051 & $-4.509\times 10^{-5}N^2 + 0.015N + 0.735Z^{1/3} + 1.132$ \\[1mm]    
\hline
\end{tabular}}
\label{tab:lgbmfold}
\end{table*}

\begin{table*}[!htb]
\centering
\caption{Expressions derived in each fold of symbolic regression using GPR numerical results. }
\resizebox{0.95\textwidth}{!}{\begin{tabular}{|c|c|c|}
\hline
Fold No. & RMSE & Expression \\  [1mm]
\hline
1 & 0.059 & $0.454A^{1/3} + 0.628Z^{1/3} - 0.054(0.325CF - 1.141)((1.913\times 10^{-6}BEA - 2.105)(1.351Z^{1/3} - 5.031) - 1.250) + 0.144$ \\[1mm]
2 & 0.051 & $-0.899(0.016(0.319CF - 1.136)^2 - 0.771)(6.168\times 10^{-7}BEA + 0.700I + 1.337Z^{1/3} - 5.775) + 4.750$ \\ [1mm]
3 & 0.048 & $-0.890(0.027(1.902\times 10^{-6}BEA - 2.133)(0.314CF - 1.138) - 0.491)(0.973A^{1/3} + 2.701\times 10^{-7}BEA + 0.045CF + 1.354Z^{1/3} - 10.435) + 4.778$ \\ [1mm]
4 & 0.059 & $1.468\times 10^{-7}BEA + 0.696I + 1.113Z^{1/3} - 0.255(0.321CF - 1.143)(-0.976A^{1/3} + 1.917\times 10^{-6}BEA + 2.754) + 0.332$ \\ [1mm]
\hline
\end{tabular}}
\label{tab:gprfold}
\end{table*}

\begin{figure*}[!htb]
    \centering
    \includegraphics[width=0.45\textwidth]{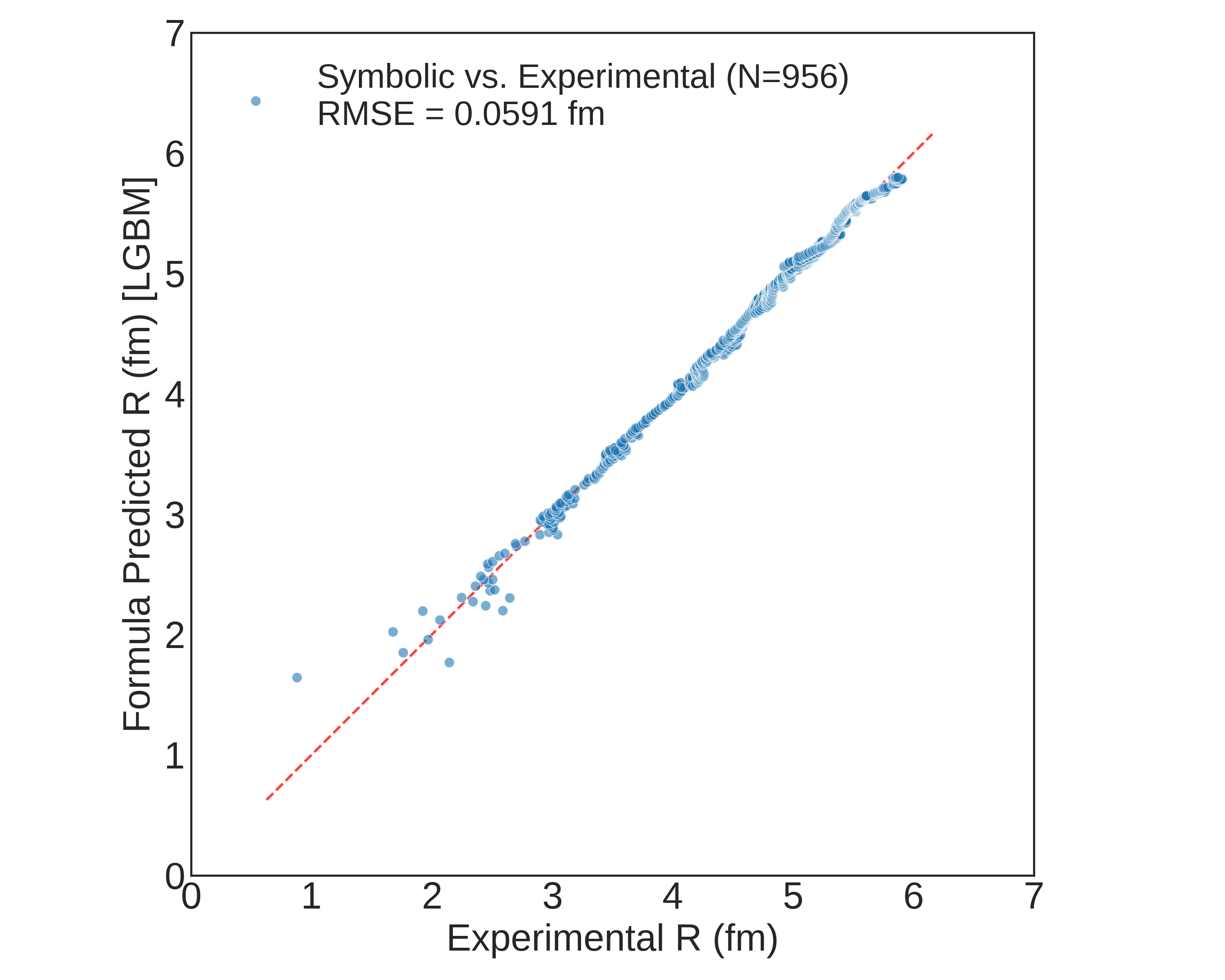}
    \includegraphics[width=0.45\textwidth]{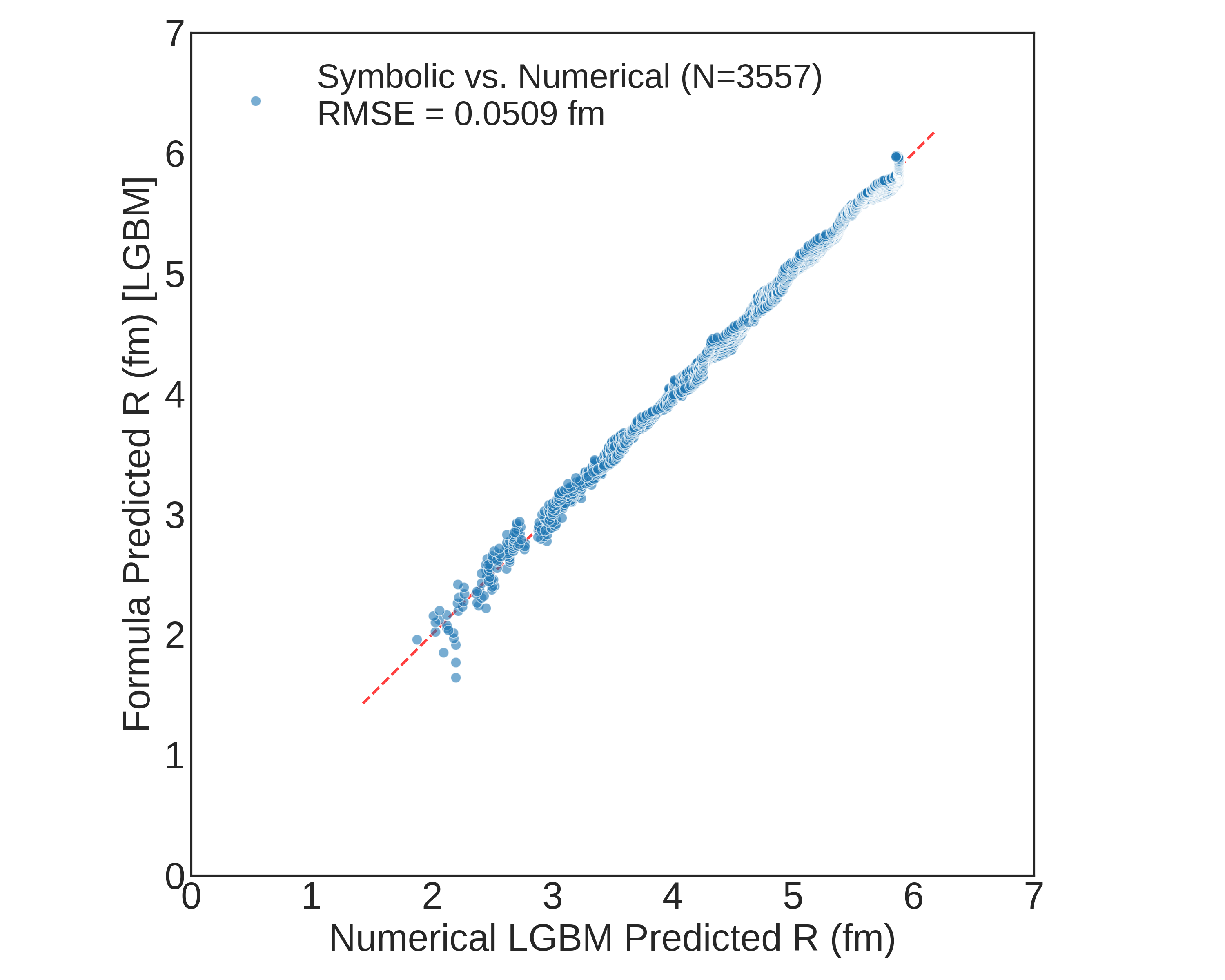}
    \includegraphics[width=0.45\textwidth]{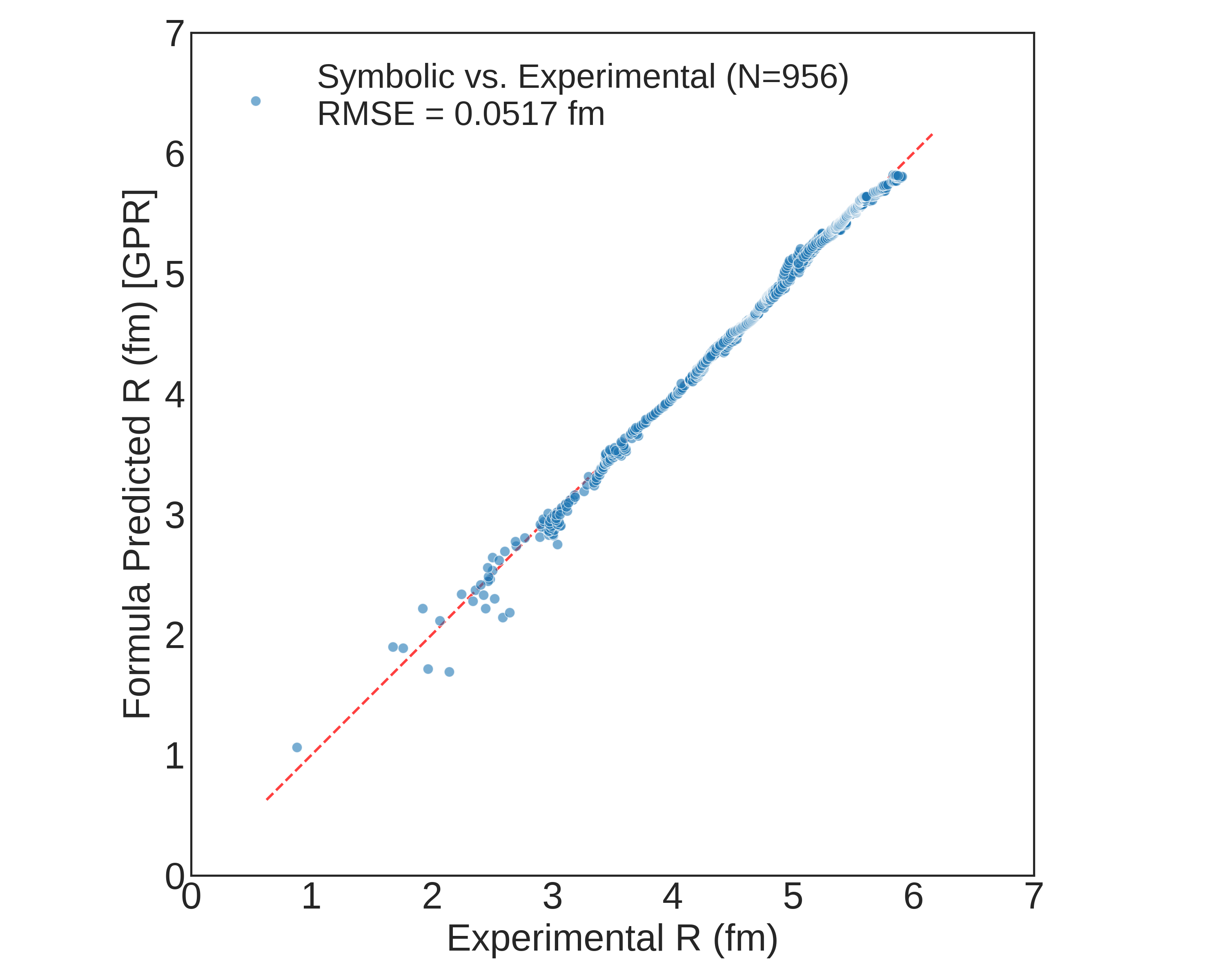}
    \includegraphics[width=0.45\textwidth]{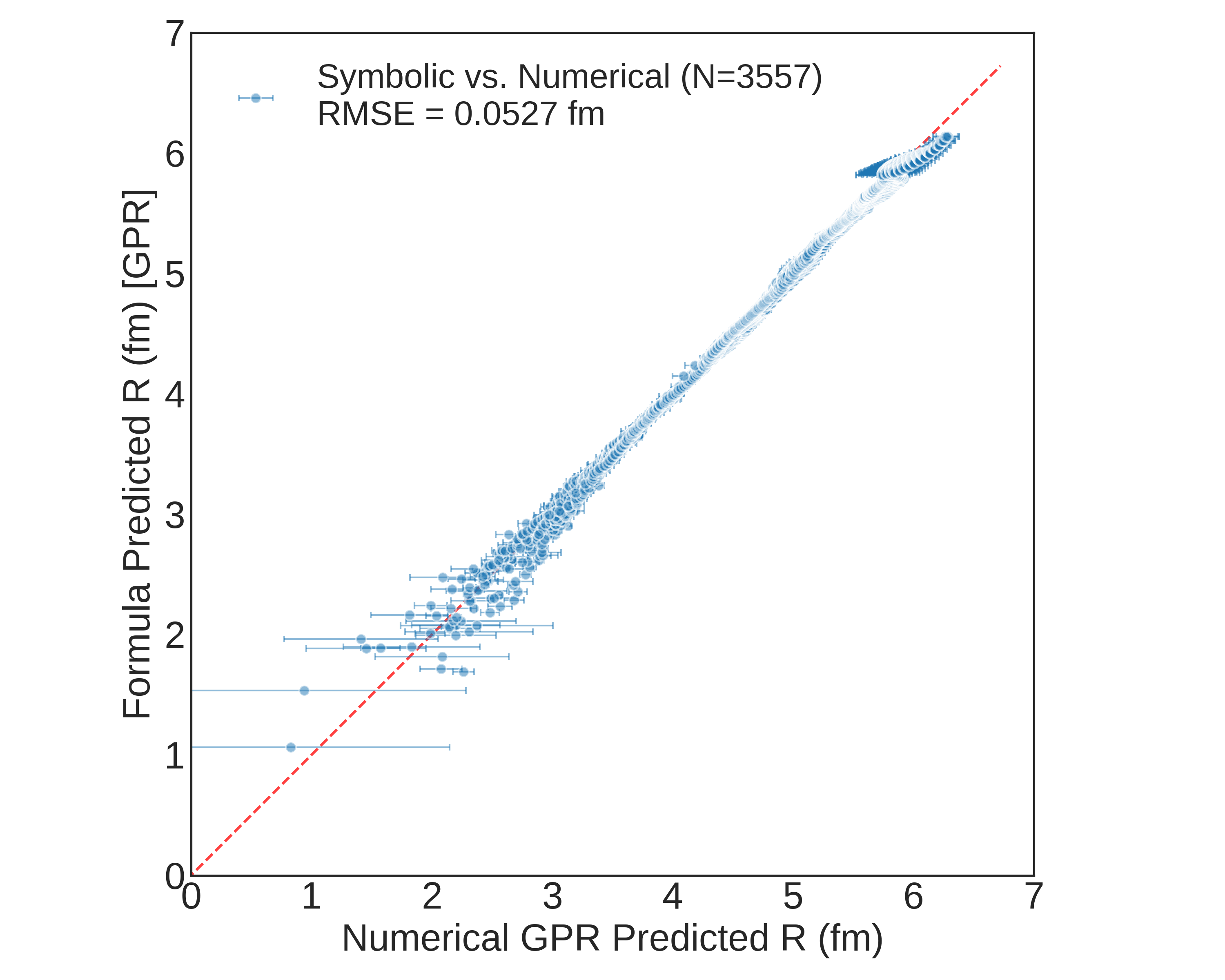}
    \caption{Comparison of LGBM and GPR formulas (at the complexity level 15) predictions with experimental data and also with numerical results.}
    \label{fig:ldata}
\end{figure*}

\begin{table*}[!htb]
\centering
\caption{Expressions derived using full data sets of LGBM results using symbolic regression. Complexity is the total count of variables, constants and operators. The algorithm skips those complexities at which it could not find an expression which was better than a simpler one it had already found.  }
\resizebox{0.95\textwidth}{!}{\begin{tabular}{|c|c|c|}
\hline
Expression & Complexity & MSE\\
\hline
$0.856 A^{1/3} + 0.471$ & 1 & 0.0100 \\[1mm]
$0.852 A^{1/3} + 0.493$ & 3 & 0.0099 \\[1mm]
$0.856 A^{1/3} - 0.472I + 0.545$ & 4 & 0.0064\\[1mm]
$0.541 A^{1/3} + 1.680\times10^{-6}BEA - 0.013N + 1.234$ & 6 & 0.0051 \\[1mm]
$0.540 A^{1/3} + 1.655\times10^{-6}BEA - 0.013N + 1.265$ & 8 & 0.0049 \\[1mm]
$-0.881(0.024(1.907\times10^{-6}BEA - 2.118)(0.023N - 1.889) - 0.552)(0.972A^{1/3} + 1.348Z^{1/3} - 9.901) + 4.756$ & 10 & 0.0042 \\[1mm]
$-0.881(0.024(0.023N - 1.889)^2 - 0.552)(0.972A^{1/3} + 1.348Z^{1/3} - 9.901) + 4.756$ & 11 & 0.0041 \\[1mm]
$-0.881(0.025(0.023N - 1.889)^2 - 0.553)(0.972A^{1/3} + 1.348Z^{1/3} - 9.910) + 4.756$ & 13 & 0.0041 \\[1mm]
$-0.881(0.032(1.907\times10^{-6}BEA - 2.118)(0.023N - 1.889) - 0.714)(0.972A^{1/3} + 0.747Z^{1/3} - 7.653) + 4.756$ & 14 & 0.0036 \\[1mm]
$-0.881(0.036(0.023N - 1.889)^2 - 0.754)(0.972A^{1/3} + 0.647Z^{1/3} - 7.281) + 4.756$ & 15 & 0.0033 \\[1mm]
\hline
\end{tabular}}
\label{tab:lgbmformulas}
\end{table*} 

\begin{table*}[!htb]
\centering
\caption{Expressions derived using full data sets of GPR results using symbolic regression. }
\resizebox{0.95\textwidth}{!}{\begin{tabular}{|c|c|c|}
\hline
Expression & Complexity & MSE\\
\hline
$0.866A^{1/3} + 0.427$ & 1 & 0.0105\\[1mm]
$0.861A^{1/3} + 0.450$ & 3 & 0.0104 \\[1mm]
$0.866A^{1/3} - 0.424I + 0.494$ & 4 & 0.0076 \\[1mm]
$0.433A^{1/3} + 0.601Z^{1/3} + 0.350$ & 5 & 0.0069 \\[1mm]
$0.522A^{1/3} + 0.477Z^{1/3} + 0.366$ & 7 & 0.0066 \\[1mm]
$0.386A^{1/3} + 1.859\times 10^{-7}BEA + 0.536Z^{1/3} + 0.617$ & 8 & 0.0063 \\[1mm]
$-0.891(0.017(0.320CF - 1.140)(1.348Z^{1/3} - 5.038) - 0.526)(0.972A^{1/3} + 1.348Z^{1/3} - 9.901) + 4.76$ & 10 & 0.0050 \\[1mm]
$0.026CF + 0.740I + 1.202Z^{1/3} + (0.056 - 0.016CF)(1.907\times 10^{-6}BEA - 2.118)^2 + 0.060$ & 12 & 0.0044 \\[1mm]
$0.027CF + 0.756I + 1.202Z^{1/3} - 0.054 (0.972 A^{1/3} - 4.862)(1.907 \times 10^{-6}BEA - 2.118) (0.320CF - 1.140) + 0.055$ & 13 & 0.0039 \\[1mm]
$0.777I + 1.202Z^{1/3} - 0.051(0.320CF - 1.140)((0.972A^{1/3} - 4.862)(1.907 \times 10^{-6}BEA - 2.118) - 1.506) + 0.148$ & 14 & 0.0038 \\[1mm]
$0.026 CF + 0.728 I + 1.202 Z^{1/3} - 0.062 (1.907\times10^{-6}BEA - 2.118)((0.972A^{1/3} - 4.862)(0.320CF - 1.140) - 0.435) + 0.063$ & 15 & 0.0035 \\[1mm]
\hline
\end{tabular}}
\label{tab:gprformulas}
\end{table*} 

Hyperparameter optimization for the composite kernel $(ConstantKernel*Matern+WhiteKernel)$ of the GPR model is performed using Optuna~\cite{optuna} for 100 trials, where $WhiteKernel$ is for observation noise. The search was performed for the constant value scaling factor $constant\_value (log, 1e-10$ to $1e5)$, characteristic $length\_scale (log, 1e-10$ to $1e5)$, both optimized on a logarithmic scale. The smoothness parameter is chosen from the discrete set $\nu ([0.5, 1.5, 2.5, 3.5, 4.5])$. Additionally, the $noise\_level (log, 1e-5$ to $1e5)$ of $WhiteKernel$ and global regularization term $\alpha$ are also tuned on a logarithmic scale $(1e-10$ to $1e5)$. The model's internal fitting process utilized the L-BFGS-B optimizer~\cite{scikit}. 

Note that the automated hyperparameter tuning process with four-fold cross-validation~\cite{kfold} using Optuna for both models is embedded inside a pipeline. In each fold of the cross-validation within every trial, the data is imputed and scaled only using the training data for that fold. This prevents data leakage and gives us an honest measure of performance. After 100 trials for each model, the algorithm reports the best combinations of hyperparameters found for both LGBM and GPR models. Following this step, the main four-fold cross-validation is performed using the best hyperparameters. In each fold, it trains over $75\%$ data and predicts the remaining $25\%$ data. These out-of-fold predictions are stored after each fold and once all four folds are complete, we have a full data set of out-of-fold predictions. These results are used to further evaluate the performance of the trained ML models and are interpreted using SHAP~\cite{shap} and a correlation analysis. Last but not least, the trained ML models on the full data set are utilized to extrapolate to the charge radii of nuclei where experimental data is missing in the data set. 

We further perform symbolic regression using PySRRegressor~\cite{pysr} to find the mathematical expressions deterministically~\cite{Poli2008} that accurately approximate the results of the trained LGBM and GPR models. The analysis is performed on the standardized results using the same standardized features as the original numerical ML models. A strategic choice is the use of maximum complexity size limit of 15 and parsimony=0.001, actively forcing the algorithm to prioritize simpler, less complex formulas, which is the main goal of a distillation from the numerically regressed ML modeling. Note that a symbolic regression search of mutating expressions is not purely bottom-up: highly accurate expressions at a high complexity often serve as parents for lower-complexity expressions that replace the complicated ones. 

We employ a rigorous four-fold cross-validation framework to ensure the robustness of discovered expressions. We also provide in Tables~\ref{tab:lgbmfold} and~\ref{tab:gprfold} the best expression obtained at or just below complexity level 15 in each fold during the cross-validation before performing symbolic regression on the full dataset. The average performance metrics, RMSE, is based on the out-of-fold predictions for each fold. Interestingly, we get four different expressions in each fold, though the dominance of $A^{1/3}$, or $Z^{1/3}$ is visible. This suggests that the relationship between nuclear features and charge radii is so complex that different mathematical expressions can achieve a similar level of accuracy. These expressions are not just the answers but the clues into the numerically regressed ML results. To synthesize the diverse insights from the four-fold cross-validation into a single, comprehensive model, the symbolic regression is ultimately trained on the full data set of 3557 nuclei to produce a generalized expression capturing the most consistent physical trends. Note that we scale all input features and target variable before training for better numerical stability and once a formula is found, it is unscaled to be in original data units. 

For each fold and for the final symbolic model training on the full dataset, the regression is configured with 1000 generations of evolutionary search, evolving for 50 distinct populations of expressions. The binary operators of addition, subtraction, multiplication, and division are allowed. The complexity of an expression is calculated as the sum of its operators, constants and variables. The maximum complexity for any given expression is strictly limited to 15 with a maximum nested depth of operators limited to 5 to avoid convoluted and deeply nested terms. Increasing the complexity level may lead to the incorporation of local structure effects such as shell gaps and odd-even staggering but may cost in terms of overfitting complex relationships. Corrective features $CF, BEA, I, P$ are given a complexity of 0.5, encouraging the algorithm to explore formulas incorporating these effects with more elaborate combinations of the fundamentally dominating ones $Z^{1/3},A^{1/3},N$ which are assigned a standard complexity of 1.0. The algorithm is designed to minimize the mean-squared error (MSE). We also list in Tables~\ref{tab:lgbmformulas} and~\ref{tab:gprformulas} the evolving history of expressions for the final symbolic model for both LGBM and GPR.

The performance metrics RMSE for both these final expressions using full LGBM and GPR model results are $\sim0.05$ $(\sqrt{MSE})$. We compare the final formula predictions for LGBM with experimental data in Fig.~\ref{fig:ldata} and with numerical results in Fig.~\ref{fig:ldata}. A similar comparison is made for the final formula predictions for GPR with the experimental data in Fig.~\ref{fig:ldata} and with numerical results in Fig.~\ref{fig:ldata}. A reasonable agreement is witnessed in all four figures, validating the symbolic regression approach.  

The high-complexity analysis reveals a clear difference between the LGBM and GPR models. The existence of multiple terms in both formulas suggests that there is no single and simple path to replicate their predictions, but rather a diverse set of complex strategies that the numerically regressed ML models have learned to achieve a higher accuracy. This finding is perfectly in line with the known complexity and multifaceted nature of interactions in many-body nuclear physics, where no single simple approach is universally applicable. The value of these high-complexity expressions lies in the fact that these could be powerful tools of revealing how sophisticated numerical regression models learn to navigate this intricate physical landscape.

\appendix*

\begin{small}
\begin{verbatim}

\end{verbatim}   
\end{small}

\newpage